\documentclass[aps,prb,twocolumn,showpacs]{revtex4}

\usepackage{graphicx}  

\begin{document}

\title{Magneto-optics of Gd and Tb in the soft x-ray resonance regions}

\author{J.E. Prieto}
\email{jeprieto@physik.fu-berlin.de} 
\author{F. Heigl, O. Krupin, G. Kaindl}
\author{K. Starke}
\affiliation{Institut f\"ur Experimentalphysik, Freie Universit\"at Berlin, Arnimallee 14, D-14195 Berlin, Germany}

\date{\today}

\begin{abstract}

We present x-ray absorption spectra around the $3d\rightarrow4f$ and 
$4d\rightarrow4f$ 
excitation thresholds of in-plane magnetized Gd and Tb films 
measured by total electron yield using circularly 
polarized synchrotron radiation. By matching the experimental 
spectra to tabulated absorption data far below and above the 
thresholds, the imaginary parts of the complex refractive index
are determined quantitatively. The associated real parts for circularly 
polarized light propagating nearly parallel or antiparallel to the 
magnetization direction are obtained through the 
Kramers-Kronig relations. The derived magnetooptical parameters are
used to calculate soft x-ray reflectivity spectra of a magnetized Gd film
at the $3d\rightarrow4f$ threshold, which are found to compare well with 
our experimental spectra.

\end{abstract}


\maketitle

\section{Introduction}

Magneto-optical (MO) effects in the visible-light region
are widely used for analyzing magnetic materials~\cite{qib00}
and have important technological applications e.g., in 
the reading process of MO disks.\cite{MO}
They are based on the fact that left- and right-hand circularly
polarized (CP) light is reflected with different intensities from 
a magnetic material depending on the local magnetization direction. 
Although MO effects are small in the visible-light region, sensitive 
detection methods yield
enough contrast to distinguish bits of opposite magnetization and
allow the observation of domain structures in optical 
microscopy.\cite{hus98} One powerful feature of MO techniques
is their capability to monitor magnetization reversal processes in applied 
magnetic fields, which is not readily achievable with electron detection
techniques due to Lorentz forces. 
MO techniques in the visible-light region involve optical transitions 
between {\em delocalized} valence states, which renders it extremely 
difficult to spectrally separate the magnetic contributions of different 
elements in compounds and in advanced layered or nanostructured materials. 
This can be a severe limitation in analyzing
magnetic nanostructures~\cite{jfg98,mj99} or
heteromagnetic systems for information storage.\cite{MO,prince98} 
 
Element sensitivity is naturally achieved by employing optical 
transitions that involve core electrons. Large magneto-optical signals 
in the x-ray region were theoretically predicted for resonant 
scattering (XRS), i.e., when the x-ray photon energy is tuned to the 
transition energy from a core level into a partially filled shell 
that contributes to an ordered magnetic moment.\cite{ers75}
In fact, XRS has been used extensively to study the magnetic 
structure of lanthanide~\cite{mgb99} and actinide~\cite{mll99} 
materials in the hard x-ray regime. 

In the soft x-ray region, large changes in the specularly reflected 
x-ray intensity at the $L_{2,3}$ edges of transition metals
upon magnetization reversal have been detected~\cite{khj90,shp98,wbh99} 
and used in element-specific studies of heteromagnetic 
systems.\cite{cck94,tonnerre95,tsb98,icf99,kko00,hkt00,ggj01,zog02}
 
In present-day thin films and multilayers, the wavelength of soft x~rays 
is comparable to the system dimensions (typical thicknesses range from  
1 to 10~nm). It has been shown by several studies~\cite{cck94,tsb98,ggj01}
that in order to extract layer-resolved magnetization profiles from 
measured soft x-ray MO signals of layered systems, a comparison with model 
calculations~\cite{sts00} of reflected intensities is usually required. 
These are based on the Fresnel equations and need accurate values of the 
MO constants as input.

Several experimental determinations of soft x-ray MO 
constants~\cite{shp98,cic98,kok00} and reflection coefficients~\cite{mag02} 
have been reported for the $L_{2,3}$ thresholds 
of ferromagnetic transition metals, but results on rare-earth elements 
have been scarce, in spite of the fact that these are often employed 
to achieve high coercive fields in magnetic layers (e.g. in spring 
magnets~\cite{fjs98}) or large perpendicular magnetic anisotropies.\cite{nak99}
Only recently has it been demonstrated that sizeable MO signals are
obtained from lanthanide elements in the soft x-ray region
at the $N_{4,5}$ thresholds.\cite{shv01} 
In a further study, a huge Faraday rotation has been predicted for
Gd films at the photon energy corresponding to this transition, making 
use of the experimentally determined difference in the refractive 
index for oppositely magnetized material.\cite{phk02} 

It is the aim of this paper to perform a quantitative determination of
the MO constants 
for the
$4d\rightarrow4f$ and $3d\rightarrow4f$ absorption thresholds of 
ferromagnetic Gd and Tb metals.
Absorption spectra, calibrated by matching to tabulated data far from 
the resonances, provide the data basis for a determination of the 
imaginary parts of the refractive index.
The associated real parts are obtained through a Hilbert
transformation using the Kramers-Kronig (KK) relations. 
In this sense, the magneto-optical parameters derived in this work 
are quantitative; they are consistent with the tabulated values far 
from the resonances, but they have not been determined in an
absolute way.
Finally, in order to illustrate the applicability of the MO constants 
obtained in this way, we calculate soft x-ray reflectivity spectra of a 
magnetized Gd film at the $M_{4,5}$ thresholds and compare them with 
our experimental reflectivity spectra.

\section{Experimental}

Experiments at the $4d\rightarrow4f$ absorption thresholds were performed 
at the UE56 undulator beamline~\cite{UE56} 
of the Berliner Elektronenspeicherring f\"ur Synchrotronstrahlung 
(BESSY II), while those at the $3d\rightarrow4f$ thresholds 
were performed at the beamline ID12-B/HELIOS-I of the European 
Synchrotron Radiation Facility (ESRF).\cite{lah99,hsl99} 
In the UE56 experiments, the photon energy resolution
was set to about ${\rm 100~meV}$~(full width at half maximum),  
which is well below the
intrinsic width of the narrow $N_{4,5}$ pre-edge absorption
lines of Gd and Tb.\cite{sna97}
By scanning the photon energy at slow speed through a synchronized movement
of monochromator and undulator, an easy normalization of the spectra
was made possible; this also allows one to exploit the
high flux of the undulator beamline of about
${\rm 10^{14}\,photons/(s\times 100~mA\times 0.1\%\,bandwidth)}$
over a wide energy range. The degree of circular polarization at this 
Sasaki-type undulator beamline is 98$\pm$2\%.\cite{UE56}
The reflectivity measurements at the Gd $M_{4,5}$ lines were also 
performed at BESSY II. 
For the absorption measurements at the $M_{4,5}$ thresholds at ID12, 
the energy resolution was set to about 0.4~eV and the degree of 
circular polarization was 94$\pm$3\%.\cite{helios}

\begin{figure}[ht]
\center{\includegraphics*[width=8.5cm]{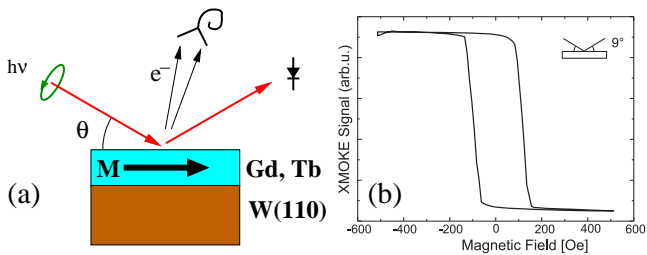}}
\caption{\label{xmoke} \small (a) Schematic experimental arrangement 
used for measuring absorption of CP light by TEY,
using a channeltron, and specular reflectivity by means of a 
photodiode. In-plane magnetized Gd and Tb epitaxial films were prepared 
on W(110) and measured {\em in situ}. (b) XMOKE hysteresis loop
of an 8~nm-thick Gd film measured at a photon energy of 147.5~eV.}
\end{figure}

Absorption spectra were recorded in total-electron yield (TEY) mode
using a high-current channeltron. To suppress the background of secondary 
electrons from the chamber walls, both the sample and a retarding grid 
placed in front of the channeltron were biased using a low-voltage battery.
For signal stability, high voltage was supplied by a 3.2-kV battery box. 
The electron-yield current was amplified by an electrometer. TEY spectra 
of the $4d\rightarrow4f$ thresholds
were normalized to the mirror current measured in the last refocusing 
element of the beamline located in a separate vacuum system in front 
of the experimental chamber.
Because it is essentially the same physical effect (secondary electron 
generation) that is used  for the measurement of both absorption and 
normalization signals, possible distortions due to a lack of proportionality 
between TEY and absorption coefficient~\cite{hsu00} in the wide 
photon energy range measured for the $N_{4,5}$ transitions 
are expected to be minimal.
The film thicknesses were always sufficiently larger than the electron's 
inelastic mean free path (IMFP) relevant for TEY measurements so that
contributions of the W substrate to the TEY signal are negligible. 
In addition, such a contribution would be structureless in the measured 
photon-energy ranges and would be removed by the procedure 
of matching the ends of the spectra to tabulated values (see below).
Reflectivity spectra were recorded in specular geometry with a photodiode 
mounted on a rotatable feedthrough; the diode current was measured by means 
of a low-noise electrometer. The experimental arrangement is shown
schematically in Fig.~\ref{xmoke}.

Epitaxial Gd and Tb metal films with thicknesses ranging from 10 to 50 nm 
were prepared {\em in situ} by vapor deposition in ultrahigh vacuum
on a W(110) single-crystal substrate. The base pressure in the chamber
was in the ${\rm 10^{-11}~mbar}$ range, rising to about 
${\rm 4 \times 10^{-10}~mbar}$ during film deposition.
Compared to transmission methods,
which give the absorption coefficient in a more direct way, TEY detection
has the advantage of allowing the use of metallic single crystals as
substrates. The growth of epitaxial films on them is well characterized.
In particular, controlled annealing of the deposited lanthanide films 
can be performed at the optimum temperatures for achieving smooth films
with homogeneous thicknesses.
(for details of film preparation, 
see Ref.~\onlinecite{sta00}). For remanent {\em in-plane} 
sample magnetization, an external field was applied along 
the [1$\bar{1}$0] direction of the substrate using a rotatable 
electromagnet.\cite{magnet} This corresponds to the easy axis 
of magnetization of the Gd and Tb films.
In the case of the magnetically harder Tb films, the sample was 
cooled from room temperature down to the measurement temperature 
(30~K) in an external magnetic field on the order of 0.1~T to achieve 
a single-domain magnetic structure with a high remanent magnetization.
This has been veryfied by means of the magneto-optical Kerr effect (MOKE) 
in the visible-light range in the laboratory and in the soft x-ray range
(XMOKE) at the beamline. As an example, Fig.~\ref{xmoke}(b)
shows an element-specific hysteresis loop of a thin Gd film measured 
by XMOKE.

\section{Results and Discussion}
\subsection{Absorption}

\begin{figure}[ht]
\includegraphics*[]{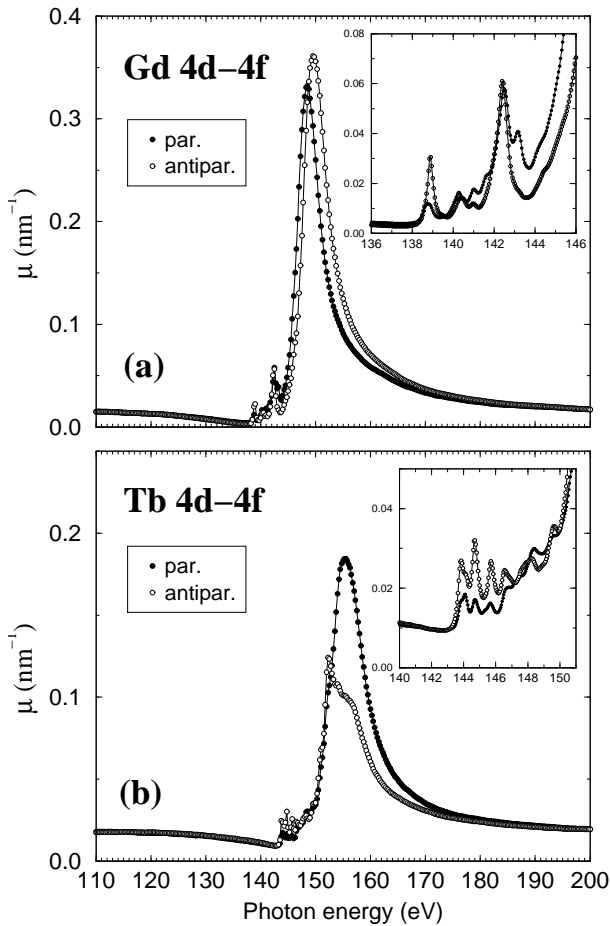}
\caption{\label{mu4dpm} \small Absorption coefficients of 
(a) Gd and (b) Tb at the $N_{4,5}$ ($4d\rightarrow4f$) thresholds of 
remanently magnetized films at ${T = 30}$~K. 
CP light was incident at $\theta{}=30^{\circ}$
with respect to the film plane, i.e. mainly parallel (filled symbols) and 
antiparallel (open symbols) to the in-plane sample magnetization.
The inserts show the regions of the prepeaks measured with
higher point densities.}
\end{figure}

\begin{figure}[ht]
\includegraphics*[]{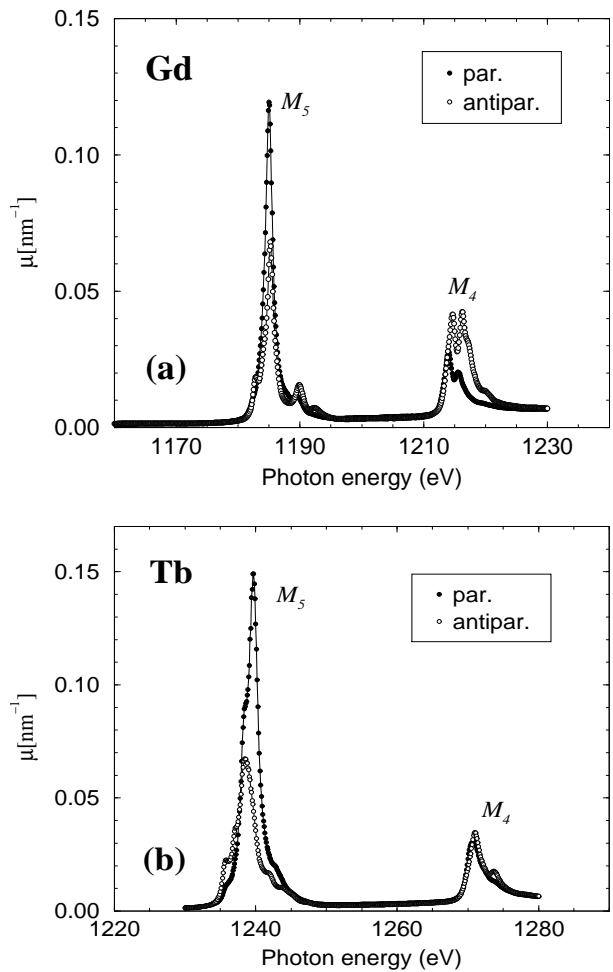}
\caption{\label{mu3dpm} \small Absorption coefficients of (a) Gd and (b) Tb
at the $M_{4,5}$ ($3d\rightarrow4f$) thresholds of
remanently magnetized films at ${T = 30~K}$.
CP light was incident at $\theta{}=30^{\circ}$
with respect to the film plane, i.e. mainly parallel (filled symbols) and
antiparallel (open symbols) to the in-plane sample magnetization.
The raw data were taken from Refs.~\cite{lah99} and~\cite{hsl99}. }
\end{figure}

Figure~\ref{mu4dpm} displays experimental absorption spectra in the 
region of the Gd and Tb $N_{4,5}$ ($4d\rightarrow4f$) thresholds for nearly 
parallel and antiparallel orientation of the magnetization with 
respect to the spin of the incoming CP photons. 
Figure~\ref{mu3dpm} shows the $M_{4,5}$ ($3d\rightarrow4f$)
absorption thresholds of both elements.
The spectra have been corrected for saturation and scaled to fit the
tabulated values of Henke {\em et al.}\cite{henke} at the ends of the
measured photon energy ranges following a procedure that is described 
in detail below.

The measured TEY spectra are affected by intrinsic saturation, which 
becomes significant when the attenuation length of the electromagnetic 
radiation $\mu^{-1}$ gets comparable to $d_e$, the inelastic mean 
free path (IMFP) of electrons in the solid.\cite{nsi99,vt88}
In that case, the detected signal at the channeltron is no longer
proportional to the absorption coefficient $\mu$. Due to the large 
absorption cross sections at the lanthanide $4d\rightarrow4f$ 
and $3d\rightarrow4f$ thresholds, the intrinsic saturation effect 
is significant and the TEY spectra must be corrected
for it in order to achieve a reliable quantitative estimation of the
absorption coefficient. 
The saturation correction was made by inverting the relation~\cite{vt88}
\begin{eqnarray} \label{eq:satcorr}
Y = C \frac{\mu d_e}{\mu d_e + \sin \theta}
\end{eqnarray}
for the measured yield $Y$.
Since the constant $C$ depends on geometrical parameters of the detection
system and is not precisely known, we need to determine the value of
$\mu d_e$ at some fixed photon energy from independent measurements. 
In the case of Gd $M_{4,5}$, the saturation correction is based on previous 
results obtained from the (6$\times$6) Eu/Gd(0001) surface.\cite{ask98} 
It consists of a quasi close-packed Eu monolayer on top of the Gd(0001) 
surface. 
Since divalent Eu has the same electronic 4$f$ shell configuration (4$f^7$) 
as trivalent Gd, both elements show the same 
multiplet structure in absorption at the $M_{4,5}$ thresholds.
The transitions are well separated in energy.
A value of (0.30$\pm$0.05) for the product $(\mu{}d_e)_{\rm peak}$ of Gd at 
the $M_{5}$ threshold has been determined by comparison of the experimental
$M_{5}$ peaks of Gd and Eu; for the latter, negligible saturation
can be safely assumed due to the small thickness of the Eu overlayer 
(1~atomic layer).\cite{sahunp}
Furthermore, by comparing the total (energy-integrated)
$M_{5}$ absorption signals of Gd and Tb and taking into 
account the different number of 4$f$ holes of both elements 
(seven in Gd, six in Tb), we estimate a value of (0.20$\pm$0.05) for 
$(\mu{}d_e)_{\rm peak}$ at the Tb $M_{5}$ threshold.
A consistency argument can be applied to justify the use of the same 
values for the saturation correction at the $4d\rightarrow4f$ thresholds:
the relevant electron IMFP at the $4d\rightarrow4f$ thresholds corresponds
to about 150~eV kinetic energy of the Auger electrons which trigger the
secondaries detected at the channeltron; it is typically 3~times smaller 
than at $\sim$~1200~eV, according to the ``universal'' curve,\cite{sd79}
while the values obtained for the absorption length $\mu^{-1}$ change 
roughly by the same factor (see below).

\begin{figure}[ht]
\includegraphics*[]{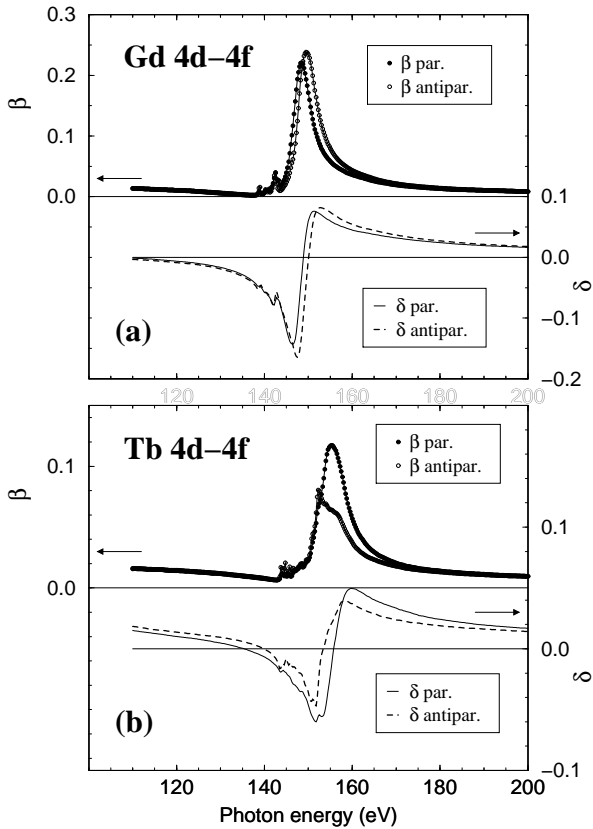}
\caption{\label{btdt4dpm} \small Optical constants ($\beta$ and $\delta$)
for (a) Gd and (b) Tb corresponding to the $N_{4,5}$ ($4d\rightarrow4f$)
thresholds of remanently in-plane magnetized films with the magnetization
vector nearly parallel (filled symbols) and antiparallel (open symbols)
to the spin of CP photons.}
\end{figure}

The saturation-corrected absorption spectra were scaled
to fit tabulated values. The photon energy ranges of the present
spectra are significantly wider than those of previous
studies;\cite{sna97,muto94} they include the wide asymmetric flanks in
the case of the ${4d\rightarrow 4f}$ resonances.  
This allows us to calibrate the absorption spectra by matching
both ends to the tabulated absorption coefficients~\cite{henke}
at photon energies where the influence of the resonances is 
expected to be negligible.
To this end we fixed the absorption coefficients $\mu_\pm$ at the low-
and high-energy sides of the measured spectra to the values
given by the tables of Henke {\em et al.}\cite{henke}
This procedure defines the ordinate scales in Figs.~\ref{mu4dpm}
and~\ref{mu3dpm}.
The values for the absorption lengths $\mu^{-1}$ determined 
in this way are given in Tables~\ref{4dtable} and~\ref{3dtable}.  
The error bars for the quoted values have been carefully estimated. 
They result from an experimental precision of $\pm$1\% at 
the considered energy and at both ends of the photon energy range, 
where the spectra were matched to the tabulated data.
This has been determined from the scatter of the data points about
their mean values. An additional contribution results from the 
uncertainty in the parameter $(\mu{}d_e)_{\rm peak}$ used in the 
saturation correction (see above).
A full error propagation calculation was made considering the explicit
functional relation of these quantities.

\begin{table}[ht]
\caption{Absorption lengths $\mu^{-1}$ (in~nm) for Gd and Tb at 
the $4d\rightarrow4f$ thresholds for nearly parallel and antiparallel 
orientations of (in plane) magnetization and photon spin. 
Values are given at the antiresonance (AR), the largest prepeak and 
the maximum of the giant resonance of each element.} \label{4dtable}
\begin{center}
\begin{tabular}{|cc|c|c|c|} \hline
element & magnet./spin & AR & prepeak & peak \\ \hline
Gd & par. & 250 $\pm$ 130 & 17 $\pm$ 3 & 3.0 $\pm$ 0.6 \\
 & antipar. & 360 $\pm$ 270 & 16 $\pm$ 3 & 2.8 $\pm$ 0.6 \\ \hline
Tb & par. & 105 $\pm$ 20 & 58 $\pm$ 12 & 5.4 $\pm$ 1.6 \\
 & antipar. & 106 $\pm$ 20 & 31 $\pm$ 8 & 8.1 $\pm$ 2.4 \\ \hline
\end{tabular}
\end{center}
\end{table}

\begin{figure*}[ht]
\begin{center} \includegraphics*[width=16.0cm]{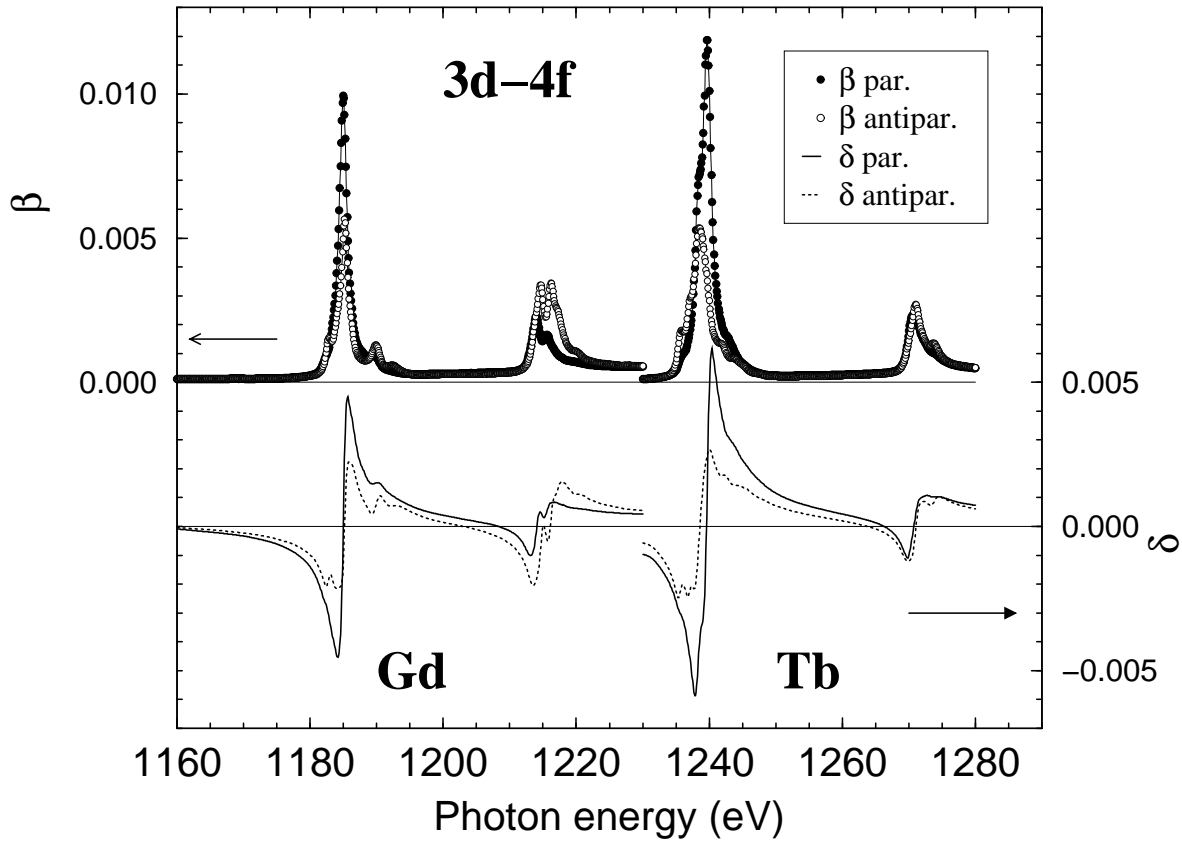}
\caption{\label{btdt3dpm} \small Optical constants ($\beta$ and $\delta$)
for Gd and Tb corresponding to the $M_{4,5}$ ($3d\rightarrow4f$)
thresholds of remanently in-plane magnetized films with the magnetization
vector nearly parallel (filled symbols) and antiparallel
(open symbols) to the spin of the CP photons.}
\end{center}
\end{figure*}

The absorption spectra at the $4d\rightarrow4f$ thresholds display
overall shapes of Fano resonances.\cite{fano61,zfg67} 
In addition, up to 10~eV below the {\em giant resonance}, 
small sharp {\em prepeaks} can be seen, which correspond
to absorption final states reached from the respective ground
states (Gd~$^8S_{7/2}$, Tb~$^7F_6$) through violation of the 
$\Delta{}L~=~0, \pm~1$ selection rule.\cite{sna97,sta00}

Note that the x-ray absorption lengths of a few nanometers at the Gd
and Tb $N_{4,5}$ maxima are the shortest in the periodic
table.\cite{henke}
They are accompannied by a huge magnetic contrast: the dichroic
(difference) signals for Gd and Tb amount respectively to 33
and 60~\% of the nonmagnetic signals, approximated by the average
of the spectral pairs in Figs.~\ref{mu4dpm} and~\ref{mu3dpm}.

The absorption cross section determined here for the Gd $N_{4,5}$
threshold is more than a factor of 2 higher than the calculated
one for atomic Eu in Ref.~\onlinecite{pan91}. These two
elements have the same 4$d^{10}$4$f^7$ ground state configuration
and hence the same 4$d^9$4$f^8$ multiplet lines, but owing to the
lower (screened)
nuclear charge of Eu one clearly expects differences in all parameters
of the giant resonance spectral profile (height, width and asymmetry).
In particular the Fano $q$~parameter depends very sensitively on the radial
matrix elements. Therefore a relative increase of the ${4d\rightarrow 4f}$
absorption maximum by a factor of 2 in going from Eu to Gd is not
unreasonable. In fact a factor of 2 is found in other 
calculations.~\cite{rmp89}

\begin{table}[ht]
\caption{Absorption lengths $\mu^{-1}$ (in~nm) for Gd and Tb at the
$3d\rightarrow4f$ thresholds for nearly parallel and antiparallel orientation
of the (in plane) magnetization and the photon spin. Values are given
at the $M_5$ and $M_4$ maxima of each element as well as at the energy
equidistant from the two maxima: 1200.9~eV for Gd and 1255.6~eV for Tb.}
\label{3dtable}
\begin{center}
\begin{tabular}{|cc|c|c|c|} \hline
element & magnet./spin & $M_5$  & betw. maxima &  $M_4$  \\ \hline
Gd & par. & 8.4 $\pm$ 1.0 & 280 $\pm$ 50 & 36 $\pm$ 4 \\
 & antipar. & 14.7 $\pm$ 1.7 & 300 $\pm$ 50 & 27  $\pm$ 3 \\ \hline
Tb & par. & 6.7 $\pm$ 1.3 & 330 $\pm$ 100 & 33 $\pm$ 7 \\
 & antipar. & 15 $\pm$ 3 & 390 $\pm$ 130 & 29 $\pm$ 6 \\ \hline
\end{tabular}
\end{center}
\end{table}

In the case of the lanthanide $M_{4,5}$ thresholds ($3d\rightarrow4f$ 
transitions), the matrix elements for Auger decay are reduced compared 
to the $4d\rightarrow4f$ resonances. This causes the Fano $q$ parameter to 
be of the order of 100.\cite{sta00} 
In the limit of large $q$, the Fano shape approaches a Lorentzian and,
in fact, the $M_{4,5}$ absorption line shape is Lorentzian, containing
hundreds of multiplet components that cluster into two main groups, 
the $M_{5}$ and $M_{4}$ thresholds\cite{gtl88} (depending on whether 
the spin of the 3$d$ hole state is oriented parallel or antiparallel to the 
$l$~=~2 orbital angular momentum). 
For several characteristic photon energies, the values of the absorption 
length, determined after the matching to tabulated data, are 
presented in Table~\ref{3dtable}.
A quantitative determination of these values seems particularly 
pertinent because standard tables~\cite{henke} do not include the 
$3d\rightarrow4f$ transitions, but only the steplike $3d\rightarrow6p$ 
threshold. 
Our results for the absorption lengths of Gd at the $M_5$ peak are 
comparable, though somewhat larger than the magnetization-averaged 
value measured in transmission in Ref.~\onlinecite{vty95}. However, 
as stated by the authors, their absolute values might be affected 
by systematic errors. Note also the huge experimental magnetic 
contrast at $M_5$, 
which leads to asymmetries of 40 and 57~\% for Gd and Tb, respectively.
Some time ago, Goedkoop {\em et al.} proposed to apply the
large magnetization dependence of lanthanide $M_{4,5}$ absorption
for constructing line filters to produce CP x~rays.\cite{gft88} 
Furthermore, the $M_{4,5}$ lines are
clearly separated in energy even for neighboring elements in the
periodic table like Gd and Tb; this represents an advantage for 
element-specific studies. 

\subsection{Magneto-optical constants}

From the values obtained for the absorption coefficient we are able
to calculate the magneto-optical constants, i.e., the real and imaginary
parts of the complex index of refraction, defined as  
\begin{eqnarray} \label{eq:ndef}
n_{\pm}(E) = 1 - \delta_{\pm}(E) -i \beta_{\pm}(E),
\end{eqnarray}
where the $+$ and $-$ signs refer to the magnetization pointing 
either parallel or antiparallel to the CP photon spin vector, respectively. 
The imaginary part is directly related to the absorption coefficient
through

\begin{eqnarray} \label{eq:beta_mu}
\beta_{\pm}(E) = \frac{1}{4\pi} \frac{hc}{E} \mu_{\pm}(E).
\end{eqnarray}

The real parts are calculated by means of a Hilbert transformation 
using the KK relations. Due to the broken time reversal 
symmetry inside a magnetized medium, the complex refractive index 
satisfies the symmetry relation 
$n_{\pm}(-E^*) = n^*_{\mp}(E)$, and one 
has to apply the modified KK equations~\cite{pva99}
\begin{eqnarray} \label{eq:KKmag}
\delta_+(E)+\delta_-(E) & = & - \frac{2}{\pi} \int_0^{\infty} E^\prime \frac{\beta_+(E^\prime)+\beta_-(E^\prime)}{(E^\prime)^2-E^2} dE^\prime,  \nonumber \\
\delta_+(E)-\delta_-(E) & = & - \frac{2E}{\pi} \int_0^{\infty} \frac{\beta_+(E^\prime)-\beta_-(E^\prime)}{(E^\prime)^2-E^2} dE^\prime.
\end{eqnarray}
In order to perform these integrations over the largest possible energy 
range, we resumed to the tabulated values for the optical constants outside 
the measured regions in the entire range from 0 to 30~keV taken 
from the compilation 
of Henke {\em et al.}~\cite{henke} We have included the relativistic 
correction affecting the asymptotic behavior of the real part of the 
atomic scattering factor at large photon energies
\begin{equation} \label{f1asymp}
f^\prime(E) \longrightarrow Z^* = Z - (Z/82.5)^{2.37},
\end{equation}
where $Z$ is the atomic number, as given by the fit of Henke 
{\em et al.}~\cite{henke} to the tabulated values of Kissel and 
Pratt.\cite{kissel90}  
The atomic scattering factor and the optical constants are related through
\begin{eqnarray}\label{f12dtbt}
\delta_\pm{}(E) & = & \frac{ r_0 h^2 c^2 n_{at} } { 2 \pi{} E^2 } f^\prime_\pm{}(E), \nonumber \\
\beta_\pm{}(E) & = & \frac{ r_0 h^2 c^2 n_{at} } { 2 \pi{} E^2 } f^{\prime \prime}_\pm{}(E), 
\end{eqnarray} 
where $r_0$ is the classical electron radius and $n_{\rm at}$ is the 
atomic density of the element.

The calculated real parts of the refractive index of Gd and Tb at the
$N_{4,5}$ and $M_{4,5}$ absorption thresholds are shown in Figs.~\ref{btdt4dpm}
and~\ref{btdt3dpm} for opposite orientations of the magnetization
and the photon spin vector.
They exhibit the well-known dispersive behavior, with tails ranging
far beyond the associated imaginary parts.
While the latter give the highest magnetic contrast in the absorption maxima,
the real parts also provide magnetic contrast in regions where
the absorption is small. This allows the performance of
magnetization-dependent measurements in reflectivity in regions of
different light-penetration depths, as the reflected signal is determined
by both the real and the imaginary parts of the refractive index.

The exact values for $n_{\pm}$ can only be
obtained in an ideal experiment where the magnetization direction
and the photon spin are strictly colinear. Since this is not feasible 
for in-plane magnetized films, there is a contribution of 
transitions with $\Delta{}m_J=0$ with a weight 1/2$\sin^2\theta$
(Ref.~\onlinecite{goedkoop}) to both the experimentally obtained 
$n_{+}$ and $n_{-}$.
This amounts to only a few percent for incidence angles up to 30$^\circ$.
Contributions from the opposite magnetization, scaling as 
1/2$(1-\cos{}\theta)^2$, can be safely neglected for those angles. 
The contribution of $\Delta{}m_J=0$ transitions cancels out if only 
the difference in the MO constants $n_{+} - n_{-}$ is required, as in, 
for example, the calculation of the Faraday effect.\cite{phk02}

\subsection{Reflectivity}

\begin{figure}[ht]
\includegraphics*[width=8.0cm]{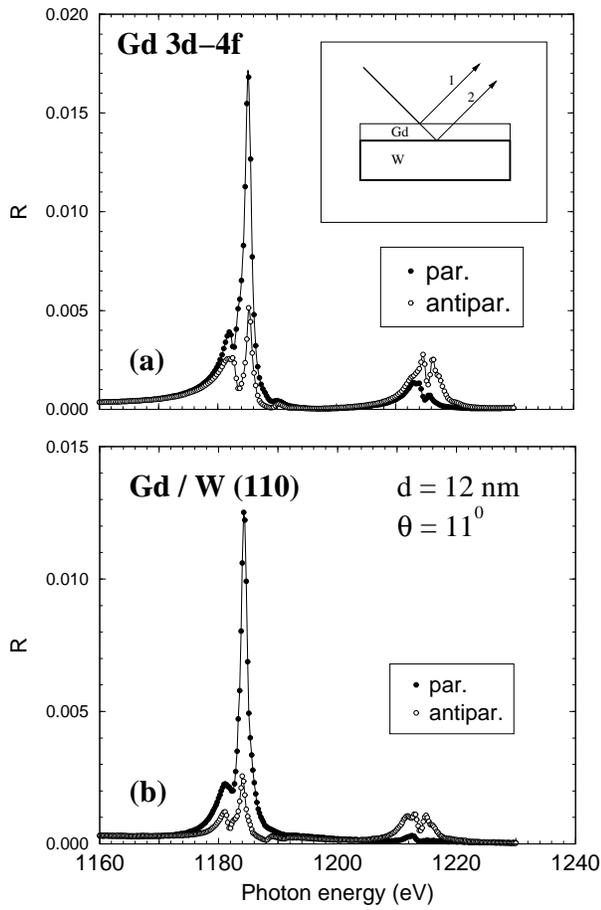}
\caption{\label{refl3dpm} \small Calculated (a) and measured (b)
x-ray reflectivity spectra in the $M_{4,5}$ region of a remanently
magnetized Gd film for nearly parallel (filled symbols) and antiparallel
(open symbols) orientation of the magnetization with respect to the light
propagation direction. Nominal film thickness is 11$\pm$1~nm and the light
incidence angle is 10$\pm{}1^{\circ}$. The values used for the calculated
spectra are 12~nm and 11$^{\circ}$.
The insert in (a) shows the two light paths considered in the calculation.}
\end{figure} 

As an application of the magneto-optical constants determined in this way, 
we present in Fig.~\ref{refl3dpm}(a) calculated reflectivity spectra of 
a Gd film in the region of the $M_{4,5}$ threshold. 
The reflectivity was 
calculated using the Jones matrix formalism and the Fresnel equations
for transmision and reflectivity at interfaces of magnetized 
media.\cite{zvk97,opp01} 
We considered interference between two channels
[see insert in Fig.~\ref{refl3dpm}(a)]: (1) represents the reflection at the 
vacuum/Gd interface and (2) comprises transmission through this 
interface, propagation in the Gd film (which includes absorption 
and Faraday effect), reflection at the Gd/W(110) surface, 
propagation back through the film, and transmission through the 
Gd/vacuum interface.
Higher-order paths, including multiple reflections, are found to 
contribute negligibly to the reflected intensity. 
In the calculation, we employed the MO constants for Gd at the 
$M_{4,5}$ thresholds shown in Fig.~\ref{btdt3dpm}, together with
the values for the W substrate taken from Ref.~\onlinecite{henke}.

Figure~\ref{refl3dpm}(b) shows the corresponding experimental
reflectivity spectra of a Gd film on W(110). The nominal film
thickness is $d=11\pm$1~nm and the light
incidence angle is $\theta{}=10\pm{}1^{\circ}$.
The experimental spectra are well reproduced by the calculation
shown in Fig.~\ref{refl3dpm}(a), including their fine structure.
The best agreement between calculated and experimental spectra was
achieved by setting $d=12~{\rm nm}$ and $\theta{}=11^{\circ}$ in the
calculation, in good agreement with the experimental values.
Quantification of the experimental reflectivity R was done by 
normalizing using the diode signal in the direct beam.
Considering the uncertainties in detector position and the
simplicity of the calculation (for example, no roughness in the 
film was considered) the agreement of the calculated absolute 
intensities with the measured ones can be considered satisfactory.

In conclusion, by quantifying magnetiza\-tion-de\-pen\-dent absorption
spectra of Gd and Tb, we were able to obtain the values of the MO
constants of both elements at the $4d\rightarrow4f$
and $3d\rightarrow4f$ excitation thresholds in the soft x-ray region.
For the example of Gd $3d\rightarrow4f$, we have shown that
the x-ray MO constants obtained here can serve as input
for reflectivity model calculations of thin lanthanide films.
In this way, our results open the
possiblity to apply the power of soft x-ray reflectivity to
element-specific studies of nanoscaled layered magnetic systems
containing lanthanide elements.

\acknowledgments
J. E. P. thanks the Alexander-von-Humboldt Stiftung for generous 
support. The authors gratefully acknowledge the experimental help
of Fred Senf and Rolf Follath (BESSY), and useful discussions with
Jeff Kortright and Eric Gullikson (LBNL). This work was 
financially supported by the German Bundesministerium f\"ur Bildung 
und Forschung, Contract No. 05 KS1 KEC/2.

\bibliography{text}

\end{document}